\documentclass[conference]{IEEEtran}

\usepackage{graphicx}
\usepackage{amsmath}
\usepackage{mathtools}
\usepackage[makeroom]{cancel}
\usepackage{amssymb}
\usepackage{amsfonts,amsthm,bm}
\usepackage{comment}
\usepackage{tasks}
\usepackage{float}
\usepackage{cite}
\usepackage{hyperref}
\usepackage{enumitem}
\usepackage{bbold}
\usepackage{gensymb}
\usepackage{multicol}

\newtheorem{theorem}{Theorem}

\newtheorem{definition}{Definition}

\begin{document}
\title{Using Passivity Theory to Interpret the \\Dissipating Energy Flow Method}

\author{\IEEEauthorblockN{Samuel Chevalier$^1$, Petr Vorobev$^{1,2}$, Konstantin Turitsyn$^1$}
\IEEEauthorblockA{$^1$Massachusetts Institute of Technology\\
$^1$Cambridge, Massachusetts\\
$^2$Skolkovo Institute of Science and Technology\\
$^2$Moscow, Russia\\
schev, petrvoro, turitsyn@mit.edu}

\and

\IEEEauthorblockN{Bin Wang}
\IEEEauthorblockA{Texas A\&M University\\
College Station, Texas\\
binwang@tamu.edu}

\and

\IEEEauthorblockN{Slava Maslennikov}
\IEEEauthorblockA{ISO New England\\
Holyoke, USA\\
smaslennikov@iso-ne.com}}
\maketitle
\begin{abstract}
Despite wide-scale deployment of phasor measurement unit technology, locating the sources of low frequency forced oscillations in power systems is still an open research topic. The dissipating energy flow method is one source location technique which has performed remarkably well in both simulation and real time application at ISO New England. The method has several deficiencies, though, which are still poorly understood. This paper borrows the concepts of passivity and positive realness from the controls literature in order to interpret the dissipating energy flow method, pinpoint the reasons for its deficiencies, and set up a framework for improving the method. The theorems presented in this paper are then tested via simulation on a simple infinite bus power system model.
\end{abstract}
\IEEEpeerreviewmaketitle


\section{Introduction}
Low frequency Forced Oscillations (FOs) are still a pervasive problem in large scale power systems. The detrimental consequences of these oscillations range from power quality degradation to dangerous resonance amplifications~\cite{MaslennikovDEF_IJEP:2017,Vanfretti:2012,Nezam:2016}. In the right context, such amplifications could lead to the tripping of a circuit breaker and a series of cascading outages. It is widely accepted across industry and academia that the most effective method for dealing with a FO is a two step process: locate the element causing the oscillation, and then disconnect it from service~\cite{MaslennikovPES:2017}. Determining the location of the source, even at the bus level, isn't a trivial task, especially when the FO frequency coincides with a natural mode of the system.

\begin{figure}
\begin{centering}
\includegraphics[scale=0.435]{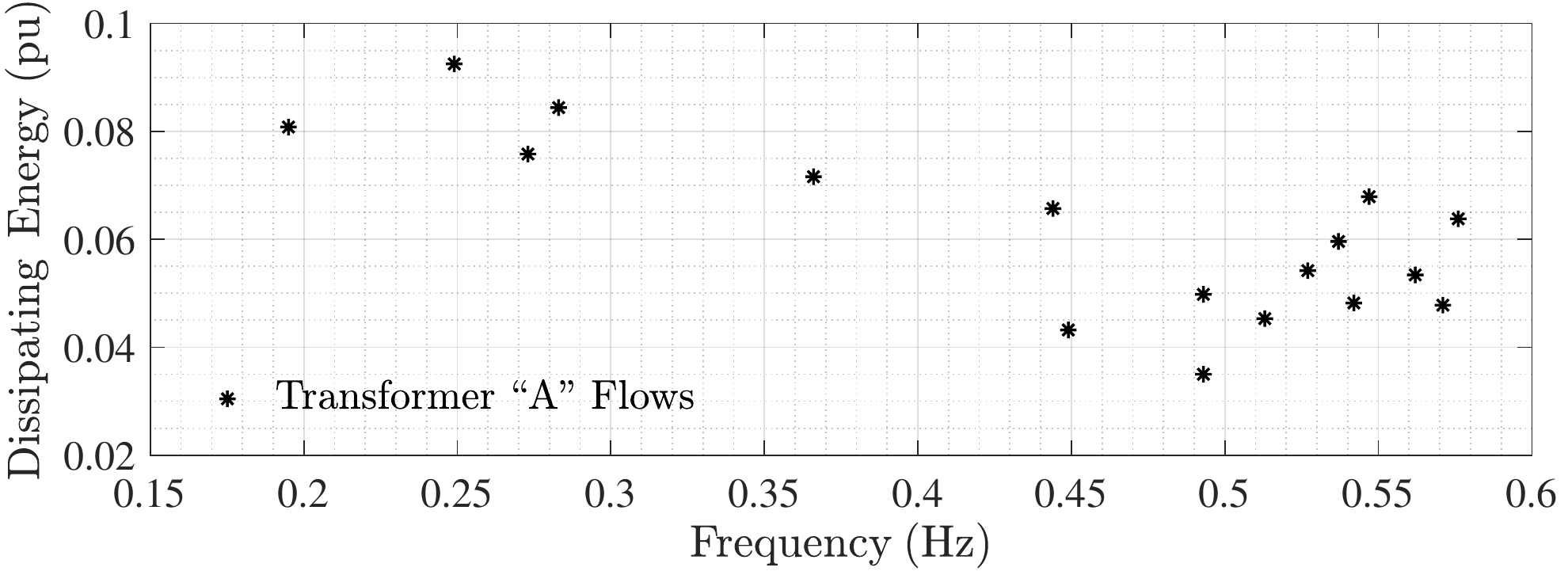}
\par\end{centering}
\caption{\label{fig: Xfm_Slava} Shown is the dissipating energy (DE) injection, flowing through transformer ``A", from a meshed load pocket near Boston, Massachusetts. 17 discrete oscillation events, spanning from 0.195 to 0.576 Hz, are presented. The source injection has been normalized to ${\rm DE}=1$.}
\end{figure}

Recently, a variety of source location algorithms have been presented \cite{Cabrera:2017,Agrawal:2017,OBrien:2017,Wilson:2014,MaslennikovDEF_IJEP:2017,Chevalier:2018,ChevalierMAP:2018,Wang:2017}. One of the most noteworthy methods is the Dissipating Energy Flow (DEF) method \cite{MaslennikovDEF_IJEP:2017} which tracks the system-wide flow  of  so-called  “transient energy”  in  order  to  locate  the source. Despite being developed in \cite{Chen:2013} under the assumptions of a lossless network and constant power loads, this  method  has  been  successfully  applied  to  many simulated test cases and hundreds of real events in the ISO New England (ISONE) system. While this method is successful in practical application, resistive transmission lines and ZIP load models with large resistances can inject positive dissipating energy (which makes them appear as sources). This phenomena has been evidenced in simulation~\cite{ChevalierMAP:2018} and in real application at ISONE. Fig. \ref{fig: Xfm_Slava}, for example, shows the dissipating energy, as quantified by~\cite[eq. (3)-(5)]{MaslennikovDEF_IJEP:2017}, injected by a meshed load pocket near Boston, Massachusetts. Shown are the injections for 17 oscillation events over the course of approximately one month in 2018. The positive injections indicate that the load is acting as a source, rather than a sink, of dissipating energy.

Explanations for the cause of resistive and conductive energy injections have been presented~\cite{Chen:2017}, but no framework has been offered which can systematically identify and fix the problem. This paper presents such a framework by employing passivity theory and defining the passivity transformation which the DEF method, implicitly, employs. Explicit conditions are derived which accurately predict when resistances will inject positive or negative dissipating energy.

The remainder of the paper is structured as follows. Section \ref{Passivity_Theory} introduces the passivity theory which is necessary for interpreting the DEF, and Section \ref{Section PSTFs} derives a series of important power system transfer functions. Section \ref{Passivity Interpretation} then analyzes the passivity of these transfer functions with the tools presented in Section \ref{Passivity_Theory}. Brief test results are provided in Section \ref{Test Results} and the paper is concluded in Section \ref{Conclusion}.

\section{Dissipative, Passive, and Positive Real Systems}\label{Passivity_Theory}
We define a dynamical system $\Sigma$ with input ${\bf u}(t)$, output ${\bf y}(t)$, and supply rate $w(t)=w({\bf u}(t),{\bf y}(t))$. In \cite{WillemsI:1972}, $\Sigma$ is defined to be dissipative if there exists a nonnegative storage function $S(\bf x)$ such that
\begin{equation}\label{eq: dissipative}
S({\bf x}_{0})+\int_{t_{0}}^{t_{1}}w(t){\rm d}t\ge S({\bf x}_{1}),
\end{equation}
where ${\bf x}_0$, ${\bf x}_1$ are the system states at times $t_0$, $t_1$ respectively. This dissipation inequality may be alternatively stated as
\begin{equation}\label{eq: dissipative_derivative}
w(t)\ge\dot{S}({\bf x}),\;\forall t
\end{equation}
which indicates that the energy supplied to the system must always be at least as large as the instantaneous change in the system's energy storage. If $w(t)=\dot{S}({\bf x}),\;\forall t$, then the system is said to be lossless and no dissipation occurs. Furthermore, $\Sigma$ is said to be \textit{passive} \cite{Bao:2007} if it is dissipative with respect to the quadratic supply rate
\begin{equation}
w\left({\bf u}(t),{\bf y}(t)\right)={\bf u}^{T}(t){\bf y}(t).
\end{equation}
If we assume $\Sigma$ is a passive system with  $S(\bf x_0)=0$ and $S(\bf x)\ge0$, then we know from (\ref{eq: dissipative}) that
\begin{equation}\label{eq: PRness}
\int_{0}^{t} {\bf u}^{T}(\tau){\bf y}(\tau){\rm d}\tau\ge0.
\end{equation}
The condition in ($\ref{eq: PRness}$) is known as positive realness, and its connections to linear systems are quite useful. Consider now linear system $\Sigma_l$ with state space representation
\begin{subequations}\label{eq: state_space}
\begin{align}
\dot{{\bf x}} & =A{\bf x}+B{\bf u}\\
{\bf y} & =C{\bf x}+D{\bf u}.
\end{align}
\end{subequations}
Assuming that this system
\begin{enumerate}[label=(\roman*)]
\item is stable, such that ${\rm Re} \left\{\lambda(A)\right\} \le 0$, and
\item does not contain repeated poles on the $j\Omega$ axis, and the residue matrix at the simple poles on the $j\Omega$ axis is Hermitian and positive semidefinite (PSD),
\end{enumerate}
then the associated transfer function
\begin{align}\label{eq: TF}
G(s)=C\left(sI-A\right)^{-1}B+D
\end{align}
is positive real if the following condition holds \cite{WillemsII:1972,slotine1991applied,khalil2002nonlinear}:
\begin{equation}
G(j\Omega) + G^T(-j\Omega)\succeq0,\;\forall \Omega\in{\mathbb R}^+,\;j\Omega\ne\lambda(A).
\end{equation}
A positive real transfer function implies passivity of the underlying system: if (\ref{eq: TF}) is positive real, $\Sigma_l$ is passive. If
\begin{equation}
G(j\Omega) + G^T(-j\Omega)\equiv{\bf 0},\;\forall \Omega\in{\mathbb R}^+,\;j\Omega\ne\lambda(A),
\end{equation}
where $\bf 0$ is the zero matrix, then $G(s)$ is said to be lossless. Notationally, we will use the Hermitian operator $\dagger$ to indicate transpose conjugation: $G^\dagger \equiv G^T(-j\Omega)$. By pushing (\ref{eq: state_space}) into the frequency domain via the Fourier transform, the state space representation may be given by
\begin{subequations}\label{eq: state_space_F}
\begin{align}
j\Omega{\tilde{\bf x}} & =A{\tilde{\bf x}}+B{\tilde{\bf u}}\\
{\tilde{\bf y}} & =C{\tilde{\bf x}}+D{\tilde{\bf u}}.
\end{align}
\end{subequations}
For this system, positive realness of (\ref{eq: TF}) corresponds to
\begin{subequations}\label{eq: Positive_scalar}
\begin{align}
{\rm Re}\left\{ \tilde{{\bf u}}^{\dagger}\tilde{{\bf y}}\right\}  & ={\rm Re}\left\{ \tilde{{\bf u}}^{\dagger}G\tilde{{\bf u}}\right\} \\
 & =\frac{1}{2}\tilde{{\bf u}}^{\dagger}\left(G+G^{\dagger}\right)\tilde{{\bf u}}\\
 & \ge0,
\end{align}
\end{subequations}
and in the time domain, (\ref{eq: PRness}) will necessarily hold. 
\begin{definition}
To be consistent with the DEF literature, we refer to $P^{\star}={\rm Re}\left\{ \tilde{{\bf u}}^{\dagger}\tilde{{\bf y}}\right\}$ as dissipating power, and we refer to the time domain integral $E^{\star}=\int_{0}^{t} {\bf u}^{T}(\tau){\bf y}(\tau){\rm d}\tau$ as dissipating energy. By the definition of positive realness,
\begin{itemize}
\item $E^{\star}>0\Leftrightarrow P^{\star}>0$
\item $E^{\star}\ge0\Leftrightarrow P^{\star}\ge0$.
\end{itemize}
\end{definition}
It it important to note that if the system 
\begin{equation}
\tilde{{\bf y}} =G\tilde{{\bf u}}
\end{equation}
isn't passive, it may be transformed into a passive system if nonsingular matrices $\bf M$ and $\Gamma$ may be identified such that
\begin{equation}
{\bf M}G{\Gamma}+({\bf M}G{\Gamma})^{\dagger}\succeq0.
\end{equation}
This corresponds to ensuring that the transformed system
\begin{equation}\label{eq: PXfm}
\left({\bf M}\tilde{{\bf y}}\right) =\left({\bf M}G{\Gamma}\right)\left({\Gamma}^{-1}\tilde{{\bf u}}\right),
\end{equation}
with input vector $\left({\Gamma}^{-1}\tilde{{\bf u}}\right)$ and output vector $\left({\bf M}\tilde{{\bf y}}\right)$, is passive. In this paper, we refer to the process of applying ${\bf M}$ and $\Gamma$ to a given FRF as a \textit{passivity transformation}.

\section{Relevant Power System Transfer Functions}\label{Section PSTFs}
Three of the dominating dynamical elements in a classical power system are constant power loads, constant impedances, and classical ($2^{\rm nd}$ order) generators. The linearized frequency response function (FRF) associated with each of these elements can be constructed for a given steady state equilibrium point. In each case, the FRF will be given in rectangular, rather than polar, coordinates: rectangular voltage perturbations will be treated as inputs and rectangular current perturbations will be treated as outputs, as in \cite{Chevalier:2018}. We first consider a constant impedance shunt element whose admittance is given by $G_z+jB_z=(R_z+jX_z)^{-1}$. If positive current is defined as flowing into the element, then Ohm's law yields
\begin{equation}\label{eq: Ohm}
I_{r}(t)+jI_{i}(t) =\left(G_z+jB_z\right)\left(V_{r}(t)+jV_{i}(t)\right).
\end{equation}
The rectangular time domain signals, such as $I_{r}(t)$, can be written as the sum of a steady state component plus a perturbation: $I_{r}(t)=I_{r} + \Delta I_{r}(t)$. For notational convenience, time dependence of the  perturbations is assumed and $\Delta I_{r}\equiv\Delta I_{r}(t)$, for example. The linear expression of (\ref{eq: Ohm}) can be expressed as a real matrix relating real vectors, where inputs and output are treated as perturbations:
\begin{equation}\label{eq: Z_Lin}
\left[\begin{array}{c}
\Delta I_{r}\\
\Delta  I_{i}
\end{array}\right] =\left[\begin{array}{cc}
G_z & -B_z\\
B_z & G_z
\end{array}\right]\left[\begin{array}{c}
\Delta V_{r}\\
\Delta V_{i}
\end{array}\right].
\end{equation}
Constant power loads are inherently nonlinear, and the real and reactive power components are given by
\begin{align}
P+jQ & =\left(V_{r}+jV_{i}\right)\left(I_{r}+jI_{i}\right)^{*}\\
P & =V_{r}I_{r}+V_{i}I_{i}\\
Q & =V_{i}I_{r}-V_{r}I_{i}.
\end{align}
By solving this set of equations for $I_r$ and $I_i$ and then linearizing around some equilibrium point, the perturbation relationships are given by
\begin{equation}\label{eq: PQ_Lin}
\left[\begin{array}{c}
\Delta I_{r}\\
\Delta I_{i}
\end{array}\right]=\left[\begin{array}{cc}
-G_p & B_p\\
B_p & G_p
\end{array}\right]
\left[\begin{array}{c}
\Delta V_{r}\\
\Delta V_{i}
\end{array}\right],
\end{equation}
where $G_p=\frac{V_{r}I_{r}-V_{i}I_{i}}{V_{r}^{2}+V_{i}^{2}}$ and $B_p=\frac{-V_{i}I_{r}-I_{i}V_{r}}{V_{r}^{2}+V_{i}^{2}}$ are defined for convenience. Since (\ref{eq: Z_Lin}) and (\ref{eq: PQ_Lin}) are linear relationships with constant coefficients, taking the Fourier transform (${\mathcal F}$) is trivial. Perturbed quantities are now treated as phasors:
\begin{align}
\left[\begin{array}{c}\label{eq: Yz}
\tilde{I}_{r}\\
\tilde{I}_{i}
\end{array}\right] & =\underbrace{\left[\begin{array}{cc}
G_{z} & -B_{z}\\
B_{z} & G_{z}
\end{array}\right]}_{\mathcal{Y}_{z}}\left[\begin{array}{c}
\tilde{V}_{r}\\
\tilde{V}_{i}
\end{array}\right]\\
\left[\begin{array}{c}
\tilde{I}_{r}\\
\tilde{I}_{i}
\end{array}\right] & =\underbrace{\left[\begin{array}{cc}
-G_{p} & B_{p}\\
B_{p} & G_{p}
\end{array}\right]}_{\mathcal{Y}_{p}}\left[\begin{array}{c}
\tilde{V}_{r}\\
\tilde{V}_{i}
\end{array}\right],\label{eq: Ypq}
\end{align}
where ${\mathcal F}(\Delta I_r(t))={\tilde I}_r(\Omega)$, for example. Again, for notational convenience, frequency dependence of the phasors is assumed and ${\tilde I}_r\equiv{\tilde I}_r(\Omega)$, for example.

The equations relating voltages and currents in a classical generator are nonlinear and differential, and the associated FRF is derived in \cite{Chevalier:2018}. For convenience, this FRF is stated:
\begin{align}
\mathcal{Y}_g &\label{eq: Yg} =\gamma\left[\!\!\begin{array}{cc}
\sin(\delta)\cos(\delta) \!&\! -\cos^{2}(\delta)\\
\sin^{2}(\delta) \!&\! -\sin(\delta)\cos(\delta)
\end{array}\!\!\right]\!+\!\left[\!\!\begin{array}{cc}
0 \!\!&\!\! \frac{1}{X_{d}'}\\
\frac{-1}{X_{d}'} \!\!&\!\! 0
\end{array}\!\!\right]\\
\gamma &=\frac{{\rm E}'^{2}}{X_{d}'^{2}}\frac{\left(M\left(j\Omega\right)^{2}+\frac{{\rm V}_{t}{\rm E}'}{X_{d}'}\cos(\varphi)\right)-j\left(\Omega D\right)}{\left(\frac{{\rm V}_{t}{\rm E}'}{X_{d}'}\cos(\varphi)-M\Omega^{2}\right)^{2}+\left(\Omega D\right)^{2}}.
\end{align}
Parameters are explained in \cite{Chevalier:2018}, but $D$ is the damping coefficient and $\delta$ is the absolute rotor angle.

\section{A Passivity Interpretation of the DEF}\label{Passivity Interpretation}
Although originally developed from an energy function perspective \cite{Chen:2013,Tsolas:1985}, the DEF method has a natural interpretation from the perspective of passivity. In this section, this interpretation is offered explicitly, and the passivity of the network elements introduced in Section \ref{Section PSTFs} are considered.

\subsection{Deriving the DEF Passivity Transformation}
The DEF is not rederived, but is instead stated in its most basic form and manipulated. The DEF integral is given by
\begin{equation}\label{eq: DEF}
W_{{\rm DE}}=\int{\rm Im}\left\{I^{*}{\rm d}V\right\}
\end{equation}
where $I$ is the complex current flowing into an element (negative injection), and ${\rm d}V$ is the corresponding complex voltage differential. This expression may be manipulated:
\begin{align}
W_{{\rm DE}} & =\int{\rm Im}\left\{\left(I_{r}-jI_{i}\right)\left({\rm d}V_{r}+j{\rm d}V_{i}\right)\right\}\\
 & =\int\left(I_{r}{\rm d}V_{i}-I_{i}{\rm d}V_{r}\right).\label{eq: W_DE_im}
\end{align}
We now write the rectangular voltage differentials as the product of time derivatives and time differentials:
\begin{align}
{\rm d}V_{i} &=\frac{{\rm d}V_{i}}{{\rm d}t}{\rm d}t\\
{\rm d}V_{r} &=\frac{{\rm d}V_{r}}{{\rm d}t}{\rm d}t.
\end{align}
The integral in (\ref{eq: W_DE_im}) may be updated via
\begin{align}
W_{{\rm DE}} & =\int\left(I_{r}\frac{{\rm d}V_{i}}{{\rm d}t}{\rm d}t-I_{i}\frac{{\rm d}V_{r}}{{\rm d}t}{\rm d}t\right)\\
 & =\int\left(I_{r}\dot{V}_{i}+(-\dot{V}_{r})I_{i}\right){\rm d}t.\label{eq: DEF_Int}
\end{align}
We now consider an associated linear system with small signal inputs $\dot{V}_{i}$, $-\dot{V}_{r}$ and outputs $I_r$, $I_i$. The dissipating energy of this system is the integral of its quadratic supply rate: 
\begin{equation}
E^{\star}=\int\left(I_{r}\dot{V}_{i}+(-\dot{V}_{r})I_{i}\right){\rm d}t.\label{eq: E_star_DEF}
\end{equation}
The FRF $\mathcal Y_d$ associated with this system satisfies
\begin{equation}\label{eq: DEF_TF}
\left[\begin{array}{c}
\tilde{I}_r\\
\tilde{I}_i
\end{array}\right]=\mathcal{Y}_d\left[\!\!\begin{array}{c}
\tilde{\dot{V}}_i\\
-\tilde{\dot{V}}_r
\end{array}\!\right]
\end{equation}
in the frequency domain. By the definition of positive realness, the integral (\ref{eq: DEF_Int}) associated with this system will be
\begin{itemize}
\item positive if $\mathcal{Y}_{d}$ is \textit{strictly positive real}:
\begin{equation}\label{eq: s_passive}
\mathcal{Y}_{d}+\mathcal{Y}_{d}^{\dagger}\succ0, \; \forall \Omega\in\mathbb R^+;
\end{equation}
\item nonnegative if $\mathcal{Y}_{d}$ is \textit{positive real}:
\begin{equation}\label{eq: p_passive}
\mathcal{Y}_{d}+\mathcal{Y}_{d}^{\dagger}\succeq0, \; \forall \Omega\in\mathbb R^+;
\end{equation}
\item and uniformly 0 if $\mathcal{Y}_{d}$ is \textit{lossless}:
\begin{equation}\label{eq: l_passive}
\mathcal{Y}_{d}+\mathcal{Y}_{d}^{\dagger}\equiv{\bf 0}, \; \forall \Omega\in\mathbb R^+.
\end{equation}
\end{itemize}
The DEF method is concerned with defining an energy function such that all elements in the power system ``dissipate" energy unless they are the source, and accordingly, the convention used in \cite{MaslennikovDEF_IJEP:2017,MaslennikovPES:2017} is for (\ref{eq: DEF_Int}) to be negative if a power system element is not the source of an oscillation. Alternatively stated, positive values of $W_{\rm DE}$ in \cite{MaslennikovDEF_IJEP:2017,MaslennikovPES:2017} indicate the injection of energy from a FO source. To be consistent with the passivity literature, this paper assumes the opposite convention: the dissipating energy integral associated with passive, non source power system elements will be positive. The following theorem provides the passivity transformation which is equivalent to the DEF integral of (\ref{eq: DEF}).
\begin{theorem}\label{theorem1}
Consider an element of a power system whose FRF $\mathcal{Y}$ satisfies
\begin{equation}\label{eq: FRF_T1}
\left[\begin{array}{c}
\tilde{I}_r\\
\tilde{I}_i
\end{array}\right]=\mathcal{Y}\left[\begin{array}{c}
\tilde{V}_r\\
\tilde{V}_i
\end{array}\right].
\end{equation}
The DEF integral of (\ref{eq: DEF}) associated with this element is guaranteed to be nonnegative if
\begin{equation}\label{eq: NSD_T1}
{\bf M}\mathcal{Y}\Gamma+\left({\bf M}\mathcal{Y}\Gamma\right)^{\dagger}\succeq0,
\end{equation}
where transformation matrices ${\bf M}$ and $\Gamma$ are defined as 
\begin{subequations}
\begin{align}
{\bf M} &=\left[\begin{array}{cc}\label{eq: M_T1}
1 & 0\\
0 & 1
\end{array}\right]\\ 
\Gamma &=\left[\begin{array}{cc}\label{eq: G_T1}
0 & -\frac{1}{j\Omega}\\
\frac{1}{j\Omega} & 0
\end{array}\right].
\end{align}
\end{subequations}
\begin{proof}
We transform (\ref{eq: FRF_T1}) via (\ref{eq: M_T1}) and (\ref{eq: G_T1}) as in (\ref{eq: PXfm}):
\begin{subequations}
\begin{align}
{\bf M}\left[\begin{array}{c}
\tilde{I}_r\\
\tilde{I}_i
\end{array}\right] & =\left({\bf M}\mathcal{Y}\Gamma\right)\Gamma^{-1}\left[\begin{array}{c}
\tilde{V}_r\\
\tilde{V}_i
\end{array}\right]\\
\left[\begin{array}{c}
\tilde{I}_r\\
\tilde{I}_i
\end{array}\right] & =\left(\mathcal{Y}\Gamma\right)\left[\begin{array}{c}
j\Omega\tilde{V}_i\\
-j\Omega\tilde{V}_r
\end{array}\right]
\end{align}
\end{subequations}
In the frequency domain, $j\Omega\tilde{V}_i=\tilde{{\dot V}}_i$ and $-j\Omega\tilde{V}_r=-\tilde{{\dot V}}_r$, so
\begin{align} \label{eq: T1_P_DEF}
\underbrace{\left[\begin{array}{c}
\tilde{I}_r\\
\tilde{I}_i
\end{array}\right]}_{\tilde{{\bf I}}} & =\left(\mathcal{Y}\Gamma\right)\underbrace{\left[\begin{array}{c}
\tilde{\dot{V}}_i\\
-\tilde{\dot{V}}_r
\end{array}\right]}_{\tilde{{\bf V}}}.
\end{align}
Positive realness of $\mathcal{Y}\Gamma$ ensures that
\begin{align}
{\rm Re}\left\{ \tilde{{\bf V}}^{\dagger}\tilde{{\bf I}}\right\}  & ={\rm Re}\left\{ \tilde{{\bf V}}^{\dagger}\left(\mathcal{Y}\Gamma\right)\tilde{{\bf V}}\right\} \ge0,\;\forall \tilde{{\bf V}}\in\mathbb{C}^{2\times1}.
\end{align}
In the time domain, this further implies that
\begin{equation}\label{eq: NRness}
\int_{0}^{t} {\bf V}^{T}(\tau){\bf I}(\tau){\rm d}\tau\ge0,
\end{equation}
where ${\mathcal F}\{{\bf V}\}= {\tilde {\bf V}}$ and ${\mathcal F}\{{\bf I}\}= {\tilde {\bf I}}$. Since the inputs and outputs of (\ref{eq: T1_P_DEF}) are identical to those of (\ref{eq: DEF_TF}), which come from the manipulated DEF integral of (\ref{eq: DEF_Int}), then positive realness of ${\bf M}\mathcal{Y}\Gamma+\left({\bf M}\mathcal{Y}\Gamma\right)^{\dagger}$ implies a positive DEF integral.
\end{proof}
\end{theorem}

\subsection{Passivity of Network Components}
We apply the results of Theorem \ref{theorem1} to consider the passivity of loads and generators. For notational simplicity, we define
\begin{equation}\label{eq: K_DEF}
K=\frac{1}{2}\left( {\bf M}\mathcal{Y}\Gamma+\left({\bf M}\mathcal{Y}\Gamma\right)^{\dagger}\right),
\end{equation}
where the passivity transformation matrices ${\bf M}$ and $\Gamma$ are given by (\ref{eq: M_T1}) and (\ref{eq: G_T1}).

\begin{theorem}\label{theorem2}
The passivity transformation from Theorem \ref{theorem1} renders linearized constant power loads lossless.
\begin{proof}
The eigenvalues of (\ref{eq: K_DEF}) for ${\mathcal Y}={\mathcal Y}_p$, from (\ref{eq: Ypq}), are:
\begin{align}
\lambda\left(K_{p}\right) & =\lambda\left(\frac{1}{2}\left(\mathcal{Y}_{p}\Gamma+\Gamma^{\dagger}\mathcal{Y}_{p}^{\dagger}\right)\right)\\
 & =\lambda\left(\frac{1}{2}\left(\mathcal{Y}_{p}\Gamma-\mathcal{Y}_{p}\Gamma\right)\right)\\
 & =\left\{ 0,\;0\right\} 
\end{align}
Since the eigenvalues are both 0, then ${\rm Re}\{ \tilde{{\bf V}}^{\dagger}\tilde{{\bf I}}\}=0$ and the constant power loads are rendered lossless.
\end{proof}
\end{theorem}

\begin{theorem}\label{theorem3}
The passivity transformation from Theorem \ref{theorem1} renders linearized classical generator models passive, but not strictly passive.
\begin{proof}
The eigenvalues of (\ref{eq: K_DEF}) for ${\mathcal Y}={\mathcal Y}_g$, from (\ref{eq: Yg}), are:
\begin{align}
\lambda\left(K_{g}\right) & =\lambda\left(\frac{1}{2}\left(\mathcal{Y}_{g}\Gamma+\Gamma^{\dagger}\mathcal{Y}_{g}^{\dagger}\right)\right)\\
 & =\left\{ -\frac{{\rm Im}\left\{ \gamma_{i}\right\} }{\Omega},\;0\right\} \\
 & =\left\{ \frac{D\frac{{\rm E}'^{2}}{X_{d}'^{2}}}{\left(\frac{{\rm V}_{t}{\rm E}'}{X_{d}'}\cos(\varphi)-M\Omega^{2}\right)^{2}+\left(\Omega D\right)^{2}},\;0\right\} \label{eq: D_gen}
\end{align}
Since one eigenvalue is strictly positive (assuming positive damping) and one eigenvalue is 0, then ${\rm Re}\{\tilde{{\bf V}}^{\dagger}\tilde{{\bf I}}\}\ge0$ and the generator is rendered passive.
\end{proof}
\end{theorem}

\begin{theorem}\label{theorem4}
The passivity transformation from Theorem \ref{theorem1} renders constant impedance loads indefinite (nonpassive) in terms of dissipating energy injection.
\begin{proof}

The eigenvalues of (\ref{eq: K_DEF}) for ${\mathcal Y}={\mathcal Y}_z$, from (\ref{eq: Yz}), are:
\begin{align}
\lambda\left(K_{z}\right) & =\lambda\left(\frac{1}{2}\left(\mathcal{Y}_{z}\Gamma+\Gamma^{\dagger}\mathcal{Y}_{z}^{\dagger}\right)\right)\\
 & =\left\{ -\frac{G_{z}}{\Omega},\;+\frac{G_{z}}{\Omega}\right\} 
\end{align}
Since the eigenvalues are equal in magnitude but opposite in sign, $K_z$ is an indefinite Hermitian matrix, and $P^{\star}={\rm Re}\{\tilde{{\bf V}}^{\dagger}\tilde{{\bf I}}\}$ can be positive or negative. Accordingly, the conductive element in a constant impedance load renders the dissipating energy injection indefinite.
\end{proof}
\end{theorem}
The results of Theorem (\ref{theorem4}) may be used to compute the conditions under which a conductance (shunt or series) will inject positive or negative dissipating energy. To do so, we manipulate the dissipating power ${{\rm Re}\{\tilde{{\bf V}}^{\dagger}\tilde{{\bf I}}\}}$:
\begin{align}
P^{\star} & =\ensuremath{{\rm Re}\left\{ \tilde{{\bf V}}^{\dagger}\left[\begin{array}{cc}
0 & -\frac{jG_{z}}{\Omega}\\
\frac{jG_{z}}{\Omega} & 0
\end{array}\right]\tilde{{\bf V}}\right\} }\\
 & =-\frac{G_{z}}{\Omega}{\rm Im}\left\{ \tilde{\dot{V}}_i^{\dagger}\tilde{\dot{V}}_r-\tilde{\dot{V}}_r^{\dagger}\tilde{\dot{V}}_i\right\}.
\end{align}
We switch to a polar representation of the voltage perturbation phasors so that $\tilde{\dot{V}}_{r}=j\Omega|\tilde{V}_{r}|e^{j\theta_{r}}$ and $\tilde{\dot{V}}_{i}=j\Omega|\tilde{V}_{i}|e^{j\theta_{i}}$ (see~\cite[Fig. 2]{MaslennikovDEF_IJEP:2017} for an interpretation):
\begin{align}
P^{\star} & = G_{z}\Omega\left|\tilde{V}_{i}\right|\left|\tilde{V}_{r}\right|{\rm Im}\left\{ e^{j\left(\theta_{r}-\theta_{i}\right)}-e^{-j\left(\theta_{r}-\theta_{i}\right)}\right\}\\
 & =2G_{z}\Omega\left|\tilde{V}_{i}\right|\left|\tilde{V}_{r}\right|\sin(\theta_{r}-\theta_{i})\label{eq: Pstar_R}
\end{align}
Therefore, the sign of the dissipating energy injection depends on $\sin(\theta_{r}-\theta_{i})$, i.e. $\sin$ of the phase shift between the input voltage perturbations. This can be shown to be consistent with the conditions given in~\cite{Chen:2017} for energy injection.

\section{Test Results}\label{Test Results}
\begin{figure}
\begin{centering}
\includegraphics[scale=0.9]{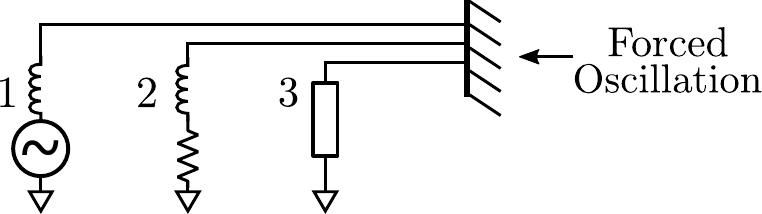}
\par\end{centering}
\caption{\label{fig: 3_Element_DEF} A generator (element 1), a constant impedance load (element 2), and a constant power load (element 3) are tied to an oscillating infinite bus.}
\end{figure}

\begin{figure}
\begin{centering}
\includegraphics[scale=0.445]{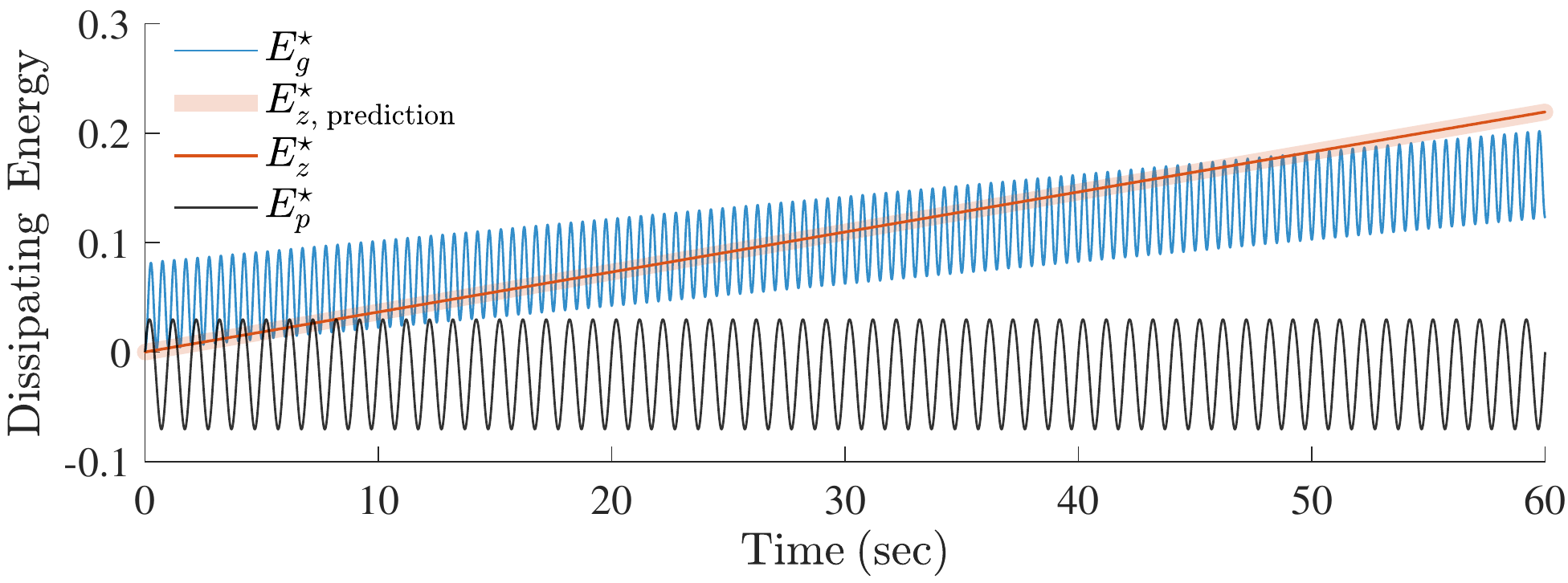}
\par\end{centering}
\caption{\label{fig: Test_1} Simulated dissipating energy flows associated with the generator, impedance, and constant power load from Fig. \ref{fig: 3_Element_DEF}. Real voltage perturbations lead the phase of imaginary voltage perturbations at the infinite bus, and generator damping is positive.}
\end{figure}

To test the results presented in Theorems (\ref{theorem1})-(\ref{theorem4}), we consider a situation in which a forced oscillation is applied by an infinite bus to the three power system elements in consideration, as depicted by Fig. \ref{fig: 3_Element_DEF}. For the sake of model explanation brevity, all simulation code has been publicly posted online\footnote{https://github.com/SamChevalier/PassiveFOs}. Two tests were run: in the first test, the phase of $\tilde{V}_{r}$ at the infinite bus \textit{led} the phase of $\tilde{V}_{i}$ by $\frac{\pi}{5}$, and the damping of the generator was assigned positive and large. The dissipating energy integrals were computed according to (\ref{eq: DEF_Int}). Additionally, the dissipating energy injected by the constant impedance load was predicted by integrating the analytical expression given in (\ref{eq: Pstar_R}). These results are shown in Fig. \ref{fig: Test_1}, where the energy dissipation is positive for the impedance load and the generator. The dissipating energy of the constant power load is lossless.

\begin{figure}
\begin{centering}
\includegraphics[scale=0.445]{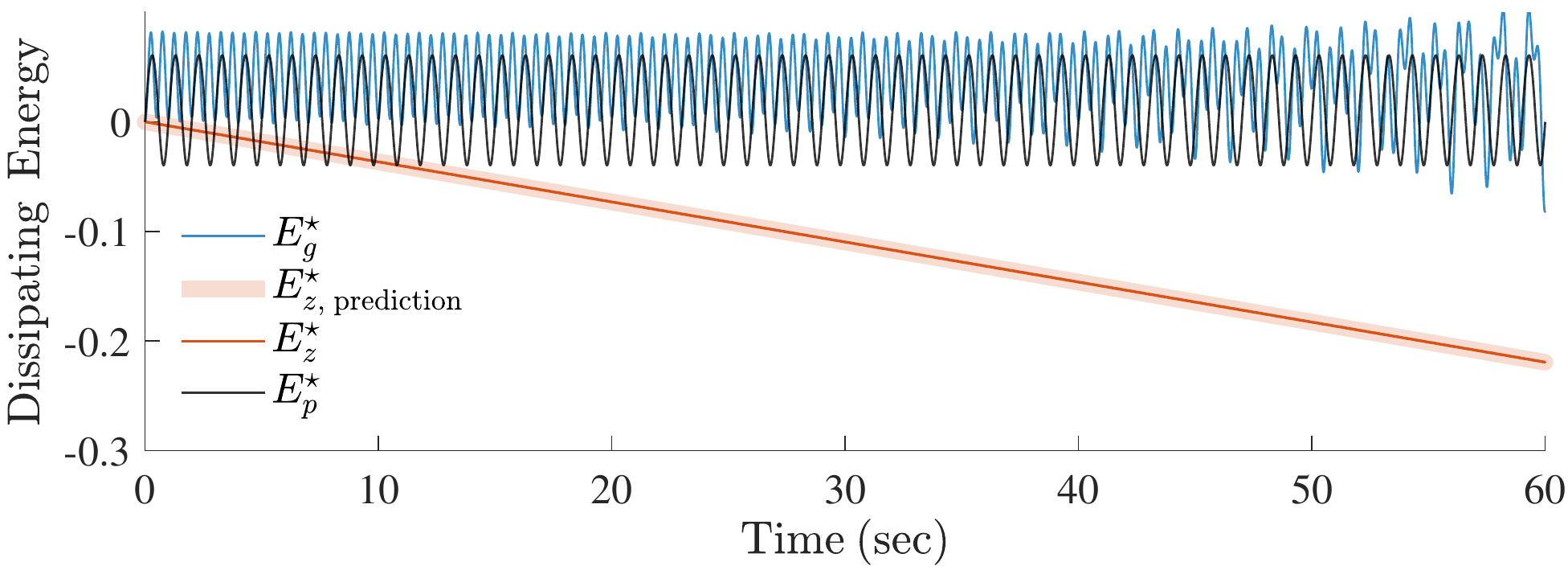}
\par\end{centering}
\caption{\label{fig: Test_2} Simulated dissipating energy flows associated with the generator, impedance, and constant power load from Fig. \ref{fig: 3_Element_DEF}. Real voltage perturbations lag the phase of imaginary voltage perturbations at the infinite bus, and generator damping is slightly negative.}
\end{figure}

In the second test, the phase of $\tilde{V}_{r}$ at the infinite bus \textit{lagged} the phase of $\tilde{V}_{i}$ by $\frac{\pi}{5}$, and the damping of the generator was assigned a slightly negative value. These results are shown in Fig. \ref{fig: Test_2}, where the energy dissipation of the impedance load has switched signs and is now negative. The constant power load is still lossless, and the generator, when stable, has a slight negative trend and therefore injects negative dissipating energy. This is predicted by the non-zero eigenvalue in (\ref{eq: D_gen}).

\section{Conclusion}\label{Conclusion}
In this paper, we interpreted the DEF method from the viewpoint of passivity. After defining the equivalent passivity transformation matrices, we showed that linearized constant power loads are lossless, linearized classical generators are passive, and constant resistance loads are indefinite in terms of dissipating energy injections. Future research will build on this analysis in order to define new transformations which improve the DEF functionality by ensuring that resistive elements don't inject negative dissipating energy.

\bibliographystyle{IEEEtran}
\bibliography{Passivity_Bib}
\end{document}